\documentclass{article}

\usepackage{arxiv}

\usepackage[utf8]{inputenc} % allow utf-8 input
\usepackage[T1]{fontenc}    % use 8-bit T1 fonts
\usepackage{hyperref}       % hyperlinks
\usepackage{url}            % simple URL typesetting
\usepackage{booktabs}       % professional-quality tables
\usepackage{amsfonts}       % blackboard math symbols
\usepackage{nicefrac}       % compact symbols for 1/2, etc.
\usepackage{microtype}      % microtypography
\usepackage{lipsum}
\usepackage{graphicx}
\graphicspath{ {./images/} }
\usepackage{algorithmic}
\usepackage{textcomp}
\usepackage{multirow}
\usepackage{hyperref}
\usepackage{amsmath}
\title{Residual Vision Transformer (ResViT) Based Self-Supervised Learning Model for Brain Tumor Classification}

\author{
\AND
Meryem Altin Karagoz \\
  Center for Diabetes Technology\\ 
  University of Virginia\\
  Charlottesville, VA 22904, USA,\\ 
  \texttt{ssy4uh@virginia.edu} \\
   \And
 O. Ufuk Nalbantoglu \\
  Department of Computer Engineering\\
  Erciyes University\\
  Kayseri,Turkey\\
  \texttt{nalbantoglu@erciyes.edu.tr} \\
  \And
Geoffrey C. Fox  \\
  Biocomplexity Institute and Initiative\\ 
  University of Virginia\\
  Charlottesville, VA 22904, USA,\\ 
  \texttt{vxj6mb@virginia.edu} \\
}

\begin{document}
\maketitle
\begin{abstract}
Deep learning has proven very promising for interpreting MRI in brain tumor diagnosis. However, deep learning models suffer from a scarcity of brain MRI datasets for effective training. Self-supervised learning (SSL) models provide data-efficient and remarkable solutions to limited dataset problems. Therefore, this paper introduces a generative SSL  model for brain tumor classification in two stages. The first stage is designed to pre-train a Residual Vision Transformer (ResViT) model for MRI synthesis as a pretext task. The second stage includes fine-tuning a ResViT-based classifier model as a downstream task. Accordingly, we aim to leverage local features via CNN and global features via ViT, employing a hybrid CNN-transformer architecture for ResViT in pretext and downstream tasks. Moreover, synthetic MRI images are utilized to balance the training set. The proposed model performs on public BraTs 2023, Figshare, and Kaggle datasets. Furthermore, we compare the proposed model with various deep learning models, including A-UNet, ResNet-9, pix2pix, pGAN for MRI synthesis, and ConvNeXtTiny, ResNet101, DenseNet12, Residual CNN, ViT for classification. According to the results, the proposed model pretraining on the MRI dataset is superior compared to the pretraining on the ImageNet dataset. Overall, the proposed model attains the highest accuracy, achieving 90.56\% on the BraTs dataset with T1 sequence, 98.53\%  on the Figshare, and 98.47\% on the Kaggle brain tumor datasets. As a result, the proposed model demonstrates a robust, effective, and successful approach to handling insufficient dataset challenges in MRI analysis by incorporating SSL, fine-tuning, data augmentation, and combining CNN and ViT.
\end{abstract}

% keywords can be removed

\section{Introduction}
Brain tumors occur due to abnormal and uncontrolled growth of cells in the brain. Brain tumors increase mortality risk and negatively affect life quality due to several medical conditions, such as hearing, vision, and sense disorders. The most prevalent types of brain tumors can be categorized into meningioma, glioma, and pituitary, with incident rates of 15\%, 45\%, and 15\%, respectively, among all brain tumors \cite{b1,b2}. The National Brain Tumor Society reports that 94,390 Americans were diagnosed with tumors in 2023, with a survival rate of 35.7\% (Brain Tumor Facts). Therefore, early diagnosis is crucial in potentially reducing mortality rate, improving treatment, implementing interventions, and improving the quality of life. For this purpose, various imaging technologies are applied to assist experts in the diagnosis of brain tumors, such as PET (positron emission tomography), MRI (magnetic resonance imaging), ultrasound screening, X-ray screening, and CT (computed tomography) \cite{b3}. MRI is a noninvasive, widely used, and effective imaging technique for early brain tumor diagnosis, enabling pain-free, high-quality 2D and 3D imaging \cite{b1}. Furthermore, MRI provides various sequences to capture different aspects of the tissues. The most commonly used sequences in brain MRI are T1-weighted, T1-weighted contrast enhancement, T2-weighted and fluid-attenuated inversion recovery (FLAIR) \cite{b4}. MRI assumes a pivotal role in the early and precise diagnosis of brain tumors owing to precise imaging, high soft tissue contrast, and comprehensive information from several sequences \cite{b5}. 

On the other hand, the examination of MRI images for brain tumor diagnosis requires an expert radiologist. The interpretation of MRI images varies subjectively between radiologists according to their experience and education. Furthermore, MRI annotation is a challenging, extensive labor, error-prone, and time-consuming process. Additionally, identifying some brain tumor cases can be challenging because of their location, characteristics, size, and visibility. Integrating artificial intelligence (AI) into computer-assisted diagnosis (CAD) systems in medical imaging has shown remarkable results in addressing these challenges. Consequently, AI-based CAD systems enhance accuracy, speed, efficiency, and consistency of diagnosis and treatment by assisting expert radiologists \cite{b1,b5,b6}.

Deep learning-based CAD models have emerged as a robust and successful tool in the automated classification of brain tumors. We present the related studies of deep learning-based models for brain tumor diagnosis via CNNs and hybrid models in the Background section. Furthermore, Table 1 presents a summary of previous studies. According to these studies, the integration of deep learning-based CAD models has significantly enhanced the efficiency and accuracy of brain tumor diagnosis, providing automated and valuable support to clinicians across various domains. However, brain tumor diagnosis with deep learning faces a shortage of annotated datasets. Publicly releasing a substantial dataset is challenging due to the need for expert annotations, privacy concerns, extensive labor, time-cost requirements, and accurate labeling \cite{b15}. The large amount of data is substantial for Deep learning models to learn a substantial number of parameters effectively and to capture diverse features. Otherwise, an insufficient dataset can lead to an over-fitting problem of deep models, thereby reducing their generalizability. Several strategies have been operated to mitigate overfitting problems such as data augmentation, synthetic data generation, multi-task learning, and transfer learning models. Developing new deep-learning strategies is still an essential subject to eliminate dataset problems, especially for medical imaging tasks.

Self-supervised learning (SSL) models have emerged as a powerful approach to addressing limited labeled data problems with prior studies explained in Section 2. Thus, self-supervised learning models provide a data-efficient approach without relying on explicit labels or large datasets by leveraging unlabeled datasets during pretext tasks \cite{b28}. From this standpoint, this study proposes a new generative self-supervised learning model for brain tumor classification on small brain MRI datasets in two stages. The proposed generative SSL model includes the pretraining stage via a Residual Vision Transformer (ResViT) for MRI synthesis as a pretext task, followed by the fine-tuning stage with a ResViT-based classifier model as a downstream task. While the pretext task network of the proposed model enables the extraction of distinct features from the self-distribution of the MRI dataset, the learned features during the pretext task are applied to the downstream classification task through fine-tuning. The proposed SSL model is constructed by combining Residual CNN and transformer blocks to leverage local and global features simultaneously for MRI image synthesis and brain tumor classification. Additionally, synthesized MRI images by ResViT are utilized as data augmentation to balance train datasets in classification. Consequently, the proposed model enables a robust, accurate, and data-efficient approach by combining various strategies to deal with overfitting problems on small brain MRI datasets: self-supervised learning,  fine-tuning, data augmentation, and hybrid architecture with CNN and ViT. The proposed model has been evaluated on Kaggle, Figshare brain tumor, and BraTs dataset for each T1, T2, and Flair sequences. This study demonstrates the comparative results of the proposed model against various deep learning models for pretext and classification tasks. Moreover, the pre-trained proposed model on BraTs has been compared with several pre-trained models on the ImageNet dataset. The Residual Vision Transformer (ResViT)-based generative self-supervised learning model provides the following contributions:
\begin{itemize}
    \item The proposed SSL model enables the extraction of local features via Residual CNN modules and global contextual features via ViT to enhance classification performances. The proposed model exhibits superior performance compared to using Residual CNN and ViT separately. Furthermore, the proposed model surpasses previous models (in Section 3.1.) for each dataset and sequence.
    \item The proposed SSL model facilitates the learning of data distribution of MRI images in an unsupervised manner (regardless of tumor type) thanks to the pretext task strategy. Implementation of transfer learning from the pre-trained network on the MRI dataset in the pretext task to the ResViT-based classifier model yields superior performance compared to state-of-the-art pre-trained models on ImageNet.
    \item This study assesses the transferability of the proposed SSL model between different MRI datasets, such as from pre-trained models on BraTs to Kaggle and Figshare Brain MRI datasets. Thus, the proposed model presents a powerful solution for processing MRI data due to its flexibility, robustness, adaptability to diverse datasets, and strong generalization ability.
    \item Utilizing both a self-learning strategy and synthesized MRI images enhances tumor classification results. 

\end{itemize}

\subsection{Background}
\subsubsection{CNN-based Deep Learning Models for Brain Tumor Diagnosis}

Deep learning-based CAD systems have demonstrated promising results for brain tumor diagnosis, particularly convolutional neural networks (CNN) \cite{b7,b8,b9}. Convolutional neural networks \cite{b10} are able to capture robust, discriminative, and local features from raw datasets owing to the convolutional mechanism \cite{b11}. Therefore, Swati \emph{et al.} \cite{b2} propose a block-wise VGG19-based brain tumor classification model via feature extraction framework and transfer learning mechanisms. Deepak \emph{et al.} \cite{b12} present a transfer learning-based brain tumor classification model by pre-trained GoogLeNet for feature extraction of brain MRI images, then fed into a classifier model. Ghassemi \emph{et al.} \cite{b13} introduce DCGAN (deep convolutional generative adversarial network) to extract robust features of MRI during the pretraining stage. The main idea of this study \cite{b13} is to leverage hierarchical representations of the brain MRI images via DCGAN. The last layer of the DCGAN replaces the fully connected layer in a classification task, followed by transfer learning and fine-tuning. Badža \emph{et al.} \cite{b14} utilize a pre-trained CNN model and data augmentation to enhance brain tumor classification performance. Ayadi \emph{et al.} \cite{b15} suggest a new CNN-based classification model for brain tumor identification on small MRI datasets. Alshayeji \emph{et al.} \cite{b16} implement optimization for hyperparameters tuning of two-path CNN. Kakarla \emph{et al.} \cite{b17} present average-pooling CNN as a lightweight model to address computational cost. Kumar \emph{et al.} \cite{b18} suggest a ResNet-50-based brain tumor classification model and utilize a global average pooling layer to eliminate overfitting problems on small datasets. Abd \emph{et al.} \cite{b19} propose BTC-fCNN for fast and efficient identification with a lightweight classification network of brain tumors. They \cite{b19} also utilize transfer learning and fine-tuning to improve classification performance. According to these studies, CNN-based brain tumor classification models are commonly utilized with transfer learning, fine-tuning, lightweight network, and average pooling strategies to avoid overfitting on a small dataset. 

\subsubsection{Hybrid Deep Learning Models via CNNs and Transformer Models for Brain Tumor Diagnosis}
Although CNNs are able to capture robust discriminative local features, CNNs have difficulties modeling long-range dependencies for medical image classification tasks \cite{b20}. The meaning of long-range in a transformer is to capture dependencies between distant tokens through self-attention mechanisms. On the other hand, CNNs are more limited in this regard, because they focus on the local area of the input operated by the filter size. Although the usage of larger filter sizes in CNNs can allow the extraction of wider features than smaller filter sizes, they still have intrinsic limitations in capturing long-range features across transformers. Medical images contain contextual relationships across both healthy and pathological tissues. Therefore, extracting robust long-range pixel features is essential for identifying medical images. For this reason, transformer models have presented an attractive and effective solution by capturing long-range feature representations owing to the self-attention mechanism. However, transformer-based models require large datasets, time, and resources to perform well \cite{b20,b21,b22,b23}. Therefore, recent studies focus on developing a hybrid model by combining CNN and transformer networks to capture both local and global features of MRI images simultaneously and reduce overfitting on small MRI datasets by enhancing the performance of brain tumor diagnosis. From this perspective, Dai \emph{et al.} \cite{b24} propose TransMed as a novel multi-modal medical image classification approach. TransMed leverages both CNN and transformer advantages to extract low-level image features efficiently and establish long-range dependencies between modalities. TransMed indicates remarkable improvements by surpassing other state-of-the-art CNN models. Aloraini \emph{et al.} \cite{b25} introduce TECNN by combining a transformer and CNN for brain MRI classification on BraTS 2018 and Figshare public dataset. TECNN obtains higher performance than both CNN-based models and ViT models. Tabatabaei \emph{et al.} \cite{b26} introduce a cross-fusion model with a parallel two-branch network that has been constructed by a transformer with a self-attention unit and lightweight CNNs for brain tumor classification. The cross-fusion model exhibits a high accuracy of 99.30\% due to combining lightweight CNNs and transformers. Ferdous \emph{et al.} \cite{b27} propose LCDEiT by utilizing a teacher-student strategy. While they use CNN as the teacher model to extract local features by reducing computational costs and dependencies of large datasets, the student model is designed by a transformer with multi-head attention mechanisms. All noted studies and their results are summarized in Table 1. All these studies indicate that utilizing CNN and transformer together has made significant progress and promising results in coping with data scarcity problems in brain tumor diagnosis. Although these methods yield highly accurate results, improving deep models is still required to deal with the lack of medical dataset problems.

\begin{table*}
    \centering
    \label{table1}
\scriptsize
\caption{The summary of previous studies regarding their publication dates, dataset utilization, model types, partitioning configurations, and accuracy results.}
\label{tab:my_label}
\begin{tabular}{|l|l|l|l|l|l|}
\hline
\textbf{References}                                                   & \textbf{Year} & \textbf{Dataset}                                                                                                                                                & \textbf{Model}                                                 & \textbf{train:val:test split} & \textbf{Accuracy}                                                                                \\ \hline
Swati \emph{et al.} \cite{b2}       & 2019          & Figshare \cite{b55}                                                                                                                            & Block-Wise VGG19                                               & 5-fold cross-val.      & 94.82                                                                                            \\ \hline
Deepak \emph{et al.} \cite{b12}     & 2019          & Figshare \cite{b55}                                                                                                                            & pre-trained GoogLeNet                                          & 5-fold cross-val.      & 97.1                                                                                             \\ \hline
Ghassemi \emph{et al.} \cite{b13}   & 2020          & Figshare \cite{b55}                                                                                                                            & DCGAN+ConvNet                                                  & 5-fold cross-val.      & 95.6                                                                                             \\ \hline
Badža \emph{et al.} \cite{b14}      & 2020          & Figshare \cite{b55}                                                                                                                            & CNN                                                            & 10-fold cross-val.     & 96.56                                                                                            \\ \hline
Ayadi \emph{et al.} \cite{b15}      & 2021          & \begin{tabular}[c]{@{}l@{}}Figshare \cite{b55} \\ Radiopaedia \\ REMBRANDT \cite{b57}\end{tabular} & CNN                                                            & 5-fold cross-val.      & \begin{tabular}[c]{@{}l@{}}Fig.: 94.74\\ Rad.: 93.71 \\ REM.: 95.72\end{tabular} \\ \hline
Alshayeji \emph{et al.} \cite{b16}  & 2021          & Figshare \cite{b55}                                                                                                                            & Aggregation of two paths CNN                                   & 70:0:30                       & 97.37                                                                                            \\ \hline
Kakarla \emph{et al.} \cite{b17}    & 2021          & Figshare \cite{b55}                                                                                                                            & average pooling+CNN                                            & 5-fold cross-val.      & 97.42                                                                                            \\ \hline
Kumar \emph{et al.} \cite{b18}      & 2021          & Figshare \cite{b55}                                                                                                                            & ResNet-50+global average pooling                               &                               5-fold cross-val.      & 97.48                                                                                            \\ \hline

Dai \emph{et al.} \cite{b24}        & 2021          & Parotid Gland Tumors                                                                                                                                     & TransMed: hybrid model via CNN and transformer            &                               80:20& 88.9                                                                                             \\\hline
Abd \emph{et al.} \cite{b19}        & 2023          & Figshare \cite{b55}                                                                                                                            & BTC-fCNN                                                       &                               5-fold cross-val.      & 98.86                                                                                            \\ \hline
Aloraini \emph{et al.} \cite{b25}   & 2023          & \begin{tabular}[c]{@{}l@{}}BraTS 2018\\ Figshare \cite{b55}\end{tabular}                                                                       & TECNN; hybrid model with transformer-enhanced CNN   & 70:20:10                      & \begin{tabular}[c]{@{}l@{}}Fig.:99.10\\ BraTS :96.75\end{tabular}                            \\ \hline
Tabatabaei \emph{et al.} \cite{b26} & 2023          & \begin{tabular}[c]{@{}l@{}}Kaggle \cite{b54} and \\ Figshare \cite{b55}\end{tabular}                                          & cross-fusion model combining CNNs and transformers & 60:20:20                      & 99.06                                                                                            \\ \hline
Ferdous \emph{et al.} \cite{b27}    & 2023          & \begin{tabular}[c]{@{}l@{}}Figshare\cite{b55} and \\ BraTS-21\end{tabular}                                                                     & LCDEiT                                                         & 10-fold cross-val.     & \begin{tabular}[c]{@{}l@{}}Fig.:98.11 \\ Brats:93.69\end{tabular}                         \\ \hline
\end{tabular}  
\end{table*}

\section{SSL Models in Medical Image Analysis}
Haghighi \emph{et al.}\cite{b29} introduce DiRA as a self-supervised learning model on unlabeled medical image datasets. DiRA consists of a discrimination component for acquiring advanced discriminative representations, a restoration component for preserving detailed information, and an adversary component for enhancing feature learning of the restoration. DiRA is evaluated on various 2D and 3D medical image datasets for classification and segmentation downstream tasks. DiRA demonstrates improvement in both classification and segmentation performance against supervised models. Furthermore, DiRA provides an efficient solution and robust model for the limited annotated medical image dataset due to collaborative self-supervised learning.

Taleb \emph{et al.} \cite{b30} propose self-supervised models through five techniques for 3D medical image segmentation and detection of downstream tasks on different medical image datasets. The pretext tasks are generated by jigsaw puzzles, contrastive predictive coding, relative patch location, rotation prediction, and exemplar networks. All of the self-supervised techniques help to learn representations of the unlabeled datasets in the first stage. Subsequently, pre-trained models in the pretext task stage are utilized for transfer learning and fine-tuning in the downstream tasks. The self-supervised strategies outperform state-of-the-art models by providing data and cost-efficient models for various medical imaging downstream tasks.

Zhou \emph{et al.}\cite{b31} introduce a self-supervised learning model, called Model Genesis, for addressing limited 3D medical image datasets. They apply four efficient image transformation techniques to learn semantic representations of the 3D medical image dataset during the image restoration task. They present a comparative study of Model Genesis against previous self-supervised and supervised models on natural image datasets for classification and segmentation. Model Genesis surpasses supervised transfer learning models on ImageNet by reducing annotation requirements of medical imaging tasks.

Srinidhi \emph{et al.} \cite{b32} offer self-supervised and semi-supervised learning models to deal with the lack of annotated dataset problems in histopathology. A self-supervised learning model is designed to capture contextual information in an unsupervised manner as a pretext task. Then, a teacher-student model is generated for consistency training via fine-tuning in classification and regression downstream tasks. Thus, their proposed model exhibits accurate, robust, and data-efficient models on limited histopathology datasets owing to the leveraging of both self and semi-supervised learning techniques.

Wang \emph{et al.} \cite{b33} present a novel self-supervised learning model by combining CNN and Swin transformer networks, called the SRCL-CTransPath, to overcome insufficient data problems in histopathological images. Initially, CTransPath is trained on large unlabeled histopathological images to extract local and global features. Then, pre-trained CTransPath performs various downstream tasks such as classification, detection, segmentation, and patch retrieval for nine public histopathology datasets. SRCL-pre-trained CTransPath model surpasses the performance of prior self-supervised models and ImageNet-pre-trained models.

Yan \emph{et al.} \cite{b34} suggest a Self-supervised Anatomical embedding model (SAM) for 2D (Xray)and 3D (CT) medical image analysis. SAM is based on a pixel-level contrastive learning and coarse-to-fine strategy to embed global and local anatomical features from unlabeled medical images. SAM outperforms well-known registration algorithms by providing fast registration. As a result, SAM exhibits a robust model with high generalization ability to apply various medical imaging modalities and medical imaging tasks such as registration, detection, and classification on small datasets. 

Kapse \emph{et al.} \cite{b35} introduce a Diversity-inducing Representation Learning (DiRL) for slide-level and patch-level histopathology classification. Digital pathology images have complex and intermixed biological components, unlike natural images. Therefore, they design pretext task for cell segmentation through a vision transformer, followed by prior and disentangle blocks, to learn context-rich representations between histopathology views. Moreover, they demonstrate the effect of attention scarification in digital pathology classification tasks. According to qualitative results, DiRL enables high achievement by capturing comprehensive and context-rich representation thanks to attention de-sparsification mechanisms.

Chen \emph{et al.} \cite{b36} present UNI as a self-supervised learning model for pathology image analysis downstream tasks. UNI utilizes DinoV2 model in the pretraining stage to extract features from the unlabeled pathology dataset. Before the downstream task, UNI is trained on a huge and diverse pathology dataset containing over 100 million tiles from various major tissue types. UNI has great potential to be used as a foundational model in anatomic pathology due to its successful outcomes and capacity to generalize and transfer across multiple tasks.

Vorontsov \emph{et al.} \cite{b37} propose Virchow as the largest model with 632 million parameters and trained 1.5 million WSIs for pathology image analysis tasks. Virchow utilizes vision transformer (ViT)-Huge and DinoV2 models in the pretraining stage on a large pathology dataset. Thus, they leverage the advantages of self-supervised learning and a teacher-student model of semi-supervised learning, as in the study of Chen \emph{et al.} \cite{b36}. Virchow is evaluated on various downstream tasks, such as prediction and classification, on different pathology datasets. Virchow gains superior and robust performance against state-of-the-art models for each dataset and downstream task owing to pretraining on large pathology image datasets.

In summary, Self-supervised learning exhibits immense potential to overcome the lack of dataset problems in various medical imaging analyses. While the SSL model emerged as a notable advancement in many medical imaging analysis tasks, there is a requirement for further progress in developing SSL models for brain tumor classification. Consequently, this study focuses on acquiring an SSL model to bridge the existing gaps and achieve more robust and effective solutions for limited dataset problems in brain tumor classification.

\section{Details of Proposed Approach}
This paper proposes a generative self-supervised learning model (SSL) for brain tumor classification in two stages. The pretext task stage includes the pretraining model before the downstream classification task for brain MRI image synthesis to learn the distribution of MRI data during synthesizing. The second stage is classification through fine-tuning from the pre-trained generative SSL model to the ResViT-based classifier model. Furthermore, synthesized MRI data are utilized as a data augmentation in the second stage to enhance classification performance. The proposed SSL model is given in Figure 1. Finally, the proposed SSL model has been compared with several state-of-the-art image synthesis and classification methods listed in Section 3.3. 
\begin{figure}[!ht]
\centerline{\includegraphics[width=0.66\columnwidth]{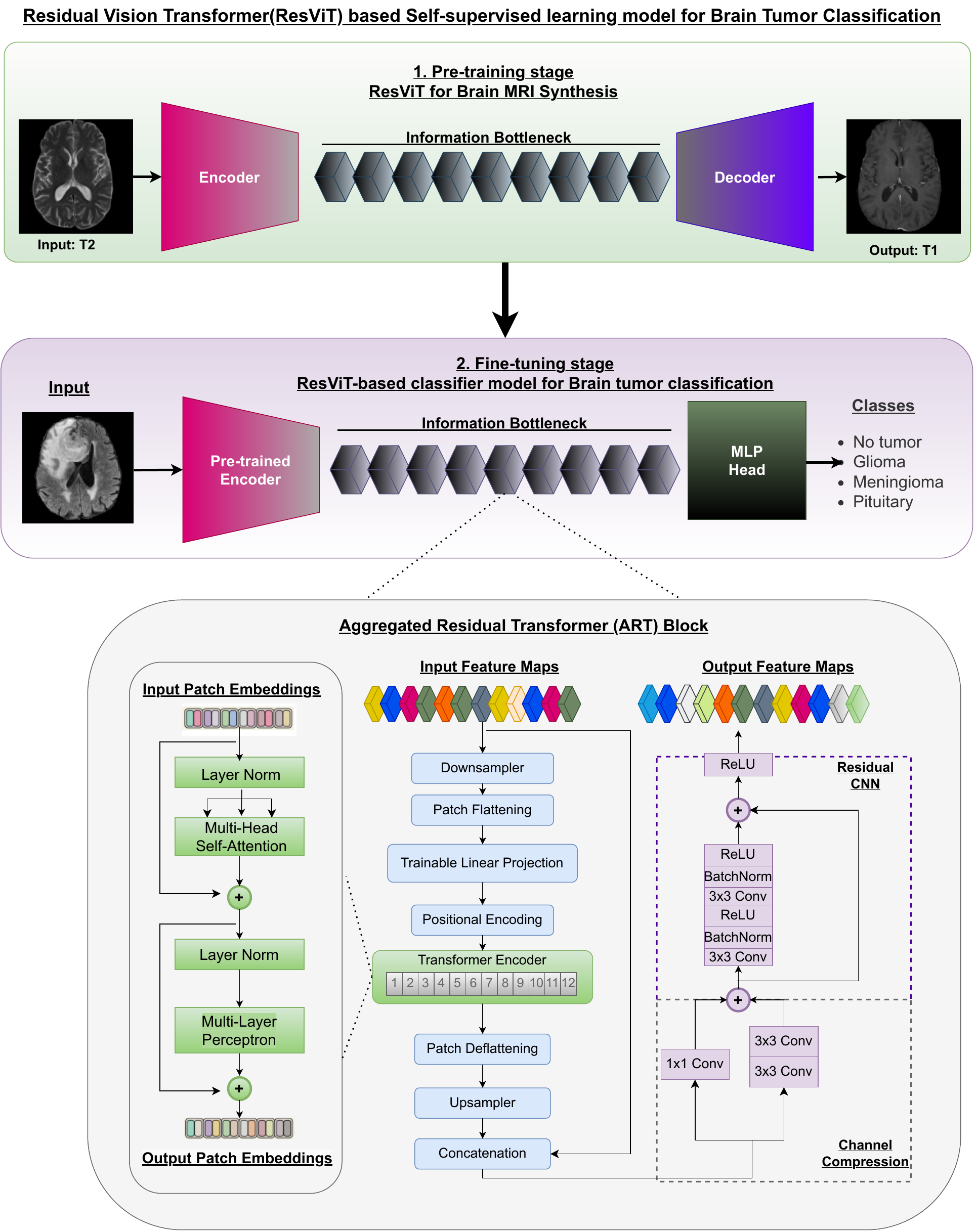}}
\caption{The proposed Residual Vision Transformer (ResViT) based Self-supervised learning model.}
\label{fig1}
\end{figure}
\subsection{Pre-training Stage: ResViT for Brain MRI Synthesis}
The first stage of the proposed model is designed for brain MRI image synthesis as a pretraining stage before the downstream task. Accordingly, the proposed model aims to capture the distribution of MRI images during the synthesized image to enhance the performance of brain tumor classification. Moreover, generated synthetic MRI data is operated as a data augmentation in the classification process. For this purpose, ResViT is utilized for brain MRI image synthesis from one sequence to the other sequence, such as from T1 to T2, T2 to T1, Flair to T1, and T1 to Flair. Residual Vision Transformers (ResViT) have been proposed by Dalmaz \emph{et al.} \cite{b38} as a generative adversarial model with hybrid architecture for medical image synthesis. ResViT combines deep residual convolutional networks and transformer blocks to capture local and global contextual features simultaneously. ResViT consists of four main subnetworks, which are the encoder, information bottleneck, decoder blocks, and discriminator. Encoder and decoder blocks are built with 2D residual CNN layers to learn the structural local features of source MRI images. On the other hand, the information bottleneck is located in the center of the network and is generated with Aggregated Residual Transformer (ART) blocks. Thus, ART blocks enable gathering information from residual CNN layers to capture structural local features and transformer branches to capture global contextual features. Finally, the discriminator is constructed by the PatchGAN \cite{b39} to ensure the generation of a realistic image by evaluating the realism of local patches in the generated (target) image compared to the corresponding patches in the real (target) image. The architecture of ResViT and ART blocks is given in Figure 1. 

The encoder maps multi-channel input X (256x256x3) onto the lower-dimensional embedded latent feature map ($f\in\ \mathbb{R}^{Nc,\ H,\ W}$), where $f,\ Nc,\ H, \ W$ are the feature map, number of channels, height, and width of the feature map, respectively. The encoded latent features are directly fed into the ART blocks via a vision transformer, the first layer of the information bottleneck. ART block consists of downsampler, patch flattening, trainable linear projection, positional encoding, the transformer encoder for patch embeddings with multi-head self-attention (MSA) \cite{b40}, and multi-layer perceptron's (MLP)\cite{b41}, patch deflattening, upsampler, concatenation layers, channel compression module and a residual CNN (ResCNN), respectively. Convolutional layers are utilized as a downsampler from $\left(f\in\ \mathbb{R}^{Nc,\ H,\ W}\right)$ to ($f^\prime\in\ \mathbb{R}^{(nc,h,w)}$) , where $nc=Nc/M,\ h=H/M,\ w=W/M,\ M$ is downsampling factor. Then, the downsampled feature map $\left(f^\prime\right)$ is divided into non-overlapping patches with a patch size of $(P, P)$and is flattened. Trainable linear projection and positional encoding are applied to the patch embeddings onto an ND-dimensional space as follows:

\begin{equation}
\mathrm{z}_\mathrm{0} = [f^{1} \, P_{E} ; f^{2} \, P_{E} ; f^{3} \, P_{E} ; \ldots ; f^{NP} \, P_{E}] + P_{E}^{\text{pos}} \in \mathbb{R}^{NP,ND}. \label{eq1}\end{equation}

where $z_{0\ }, f^p,\ P_E,\ and {\ P}_E^{pos}$ are input patch embedding, pth patch of the downsampled feature map, embedding projection, and learnable positional encoding, respectively. Following that, the transformer encoder handles patch embeddings with MSA and MLP. The output of $l^{th}$ layer belonging to the transformer encoder is given as follows, where LN is layer normalization:
\begin{equation}
z_l^\prime=MSA\left(LN\left(z_{l-1}\right)\right)+z_{l-1} .\label{eq2}\end{equation}

\begin{equation}
z_l=MLP\left(LN\left(z_l^\prime\right)\right)+z_l^\prime.\label{eq3}\end{equation}

The deflattening layer is applied for processing the upsampling layer with transposed 2D convolutions. Then, channel-wise concatenation combines the global contextual features captured by the transformer (upsampled feature map is $g$) with the local features captured by convolutional (input feature map is $f$). Channel compression is utilized for compressed channels of the concatenate feature maps $(concat(g,f))$. The final feature maps are fed into Residual CNN to obtain the output feature map of the ART block. Finally, the decoder synthesizes high-resolution MRI images from low-dimensional feature maps of the ART block via transposed convolutional layers. ResViT utilizes a linear combination of pixel-wise loss \eqref{eq5}, reconstruction loss \eqref{eq6}, and adversarial loss \eqref{eq7}, to calculate the loss function \eqref{eq8}, by the following equations.

\begin{equation}
X^G\ =a_i\cdot m_i.\label{eq4}\end{equation}

\begin{equation}
a_i=\begin{cases}
1, & \text{if } m_i \in \text{source sequence} \\
0, & \text{if } m_i \in \text{target sequence}
\end{cases}.\label{eq5}\end{equation}
\begin{equation}
L_{\text{pix}} = \sum_{i=1}^I (1 - a_i) E\left[\| (G(X^G)_i - m_i) \|_1\right].\label{eq6}\end{equation}
\begin{equation}
L_{\text{rec}} = \sum_{i=1}^I (a_i) E\left[\| G(X^G)_i - m_i \|_1\right].\label{eq7}\end{equation}

 \begin{equation} 
 L_{\text{adv}} = E\left[{D\left(X^D\left(\text{acquired}\right)\right)}^2\right]-E\left[{D\left(X^D\left(\text{synthetic}\right)-1\right)}^2\right].\label{eq8}\end{equation}

Where E is expectation, G is the generator network of ResViT, D is discriminator, $m_i (i\ \in\ \left\{1,\ 2,\ldots,\ I\right\})$ is the ith image, $X^D\left(synthetic\right)$ is the concatenation of source and synthetic images, $X^D\left(acquired\right)$ is the concatenation of the source and acquired images. Finally, the loss function of ResViT is calculated by linearly combining loss functions and their weights $(\lambda)$ in \eqref{eq9}.
\begin{equation}
L_{ResViT}=\lambda_{pix}L_{pix\ }+\ \lambda_{rec}L_{rec\ }+\lambda_{adv}L_{adv\ }.\label{eq9}\end{equation}

The encoder blocks of ResViT are constructed by stacking three convolutional layers in which the kernel size is 7,3,3 and output feature map size is 256, 64, and 64, respectively. The decoder of ResViT is the inverse of the encoder, which has three 2D transposed layers with kernel sizes 3,3,7 respectively. The information bottleneck consists of nine ART blocks $(A1, A2,\ldots, A9)$ where A1 and A6 blocks have transformer modules. While the downsampler of transformer ART blocks has two convolutional layers and is split into patches with a patch size of (16,16), the up sampler contains two 2D transposed layers for inverse processing of downsampler layers. The down sampler and upsampler are set with kernel size 3, stride 2, and downsampling factor M=4. Next, a channel compression module is applied to reduce the number of channels from 512 to 256. Furthermore, the pre-trained base (R50+ViT-B\_16, ViT-B\_16) and large \text{(ViT-L\_16)} transformer models on ImageNet \href{https://github.com/google-research/vision\_transformer} {(github.com/google-research)} are used in transformer encoder blocks. While the base model has 12 layers with 12 attention heads, dimension of latent feature (ND) =768, and 3073 hidden units, the large model has 24 layers with 16 attention heads, dimension of latent feature (ND)=1024, 4096 hidden units for each MLP layer.

\subsection{Fine-tuning stage: ResViT-based Classifier Model for Brain Tumor Classification}
The second stage of the proposed model is designed for brain tumor classification as a downstream task. The proposed ResViT-based classifier model enables end-to-end fine-tuning by transferring weight from the pre-trained ResViT to the ResViT-based classifier model and leveraging synthesized MRI images in the training phase of the classifier model. Thus, the fine-tuning stage aims to improve classification performance through self-learning. The proposed ResViT-based classifier model has three main components: the pre-trained encoder, ART blocks, and the MLP head. The proposed classifier model aims to learn local structural features via Residual CNNs and global contextual features via ViTs for brain tumor classification as in the first stage. The encoder and ART blocks of ResViT are constructed with identical setups and architecture as in the first stage, explained in the last paragraph of Section 3.1, to enable end-to-end fine-tuning. Furthermore, the decoder blocks of ResViT in the first stage are replaced with a multi-layer perceptron (MLP) for classification. The MLP head consists of a dense layer, a normalization layer, a dropout layer to reduce overfitting with a dropout rate of 0.5, and the fully connected layer with SoftMax for classifying brain tumors into "no tumor," "glioma," "meningioma," and "pituitary."
\subsection{Alternative Deep Learning Models for Comparative Study}
This study includes several state-of-the-art models for image synthesis and classification to compare with the proposed model. Attention U-Net (A-UNet) \cite{b42}, ResNet-9 \cite{b43}, pix2pix \cite{b39}, and pGAN \cite{b59} models are applied for MRI image synthesis and comparison with ResViT in the pretraining stage. The image synthesis methods are the type of conditional GAN for paired image-to-image translation. The conditional GAN models primarily consist of a generator subnetwork and a discriminator subnetwork. While A-UNet and pix2pix are constructed by Unet-based generator interconnected skip connections, ResNet-9-based conditional GAN and pGAN are generated by ResNet blocks. Furthermore, the discriminator subnetworks of the generative models have been set by the PatchGAN discriminator to assess the realism of synthesized MRI images identical to ResViT for fair comparison. Additionally, the pGAN uses pre-trained VGG16 to extract feature maps and calculate perceptual loss, unlike other models. Consequently, this study assesses the ResViT model against various state-of-the-art image-to-image translation methods.  

On the other hand, the proposed ResViT-based classifier model has been compared with ConvNeXtTiny \cite{b44}, Resnet101 \cite{b45}, and Densenet121 \cite{b46}. ConvNeXt has been introduced as a modernized version of ConvNet by adapting standard ResNet architectures into a hierarchical vision transformer. While ResNet-101 consists of residual convolutional blocks with 101 layers, DenseNet-121 is constructed as a series of dense blocks with 121 layers. The methods are pre-trained on the ImageNet dataset \cite{b47}. Thus, this study observes the effects of the proposed pre-trained and fine-tuning strategy compared to other pre-trained models on ImageNet. The proposed model combines the Residual CNN and ViT models to capture both local and global features of brain tumors. Residual CNN and ViT are individually employed for brain tumor classification to demonstrate the effectiveness of the proposed ResViT-based classifier model. Residual CNN and ViT models are constructed identically to the encoder block of ResViT and the first ART block of ResViT, respectively, for fair comparison. Furthermore, the pre-trained transformer models (R50+ViT-B\_16, ViT-B\_16, ViT-L\_16) are implemented to assess the transferability and effectiveness of leveraging pre-trained on both the natural ImageNet dataset and a dataset specific to brain MRI. In summary, we demonstrate a comprehensive comparison to observe the effectiveness of the proposed ResViT-based generative SSL model and its contributions.

\section{Experimental Study and Results}
\subsection{Dataset}
The proposed and alternative deep learning methods have been trained and evaluated on BraTS 2023 Glioma and Meningioma challenges, Kaggle brain tumor MRI dataset \cite{b54}, and Figshare brain tumor datasets \cite{b55}. BraTS is a well-known benchmark dataset that has mainly used brain tumor segmentation for 12 years. Brats 2023 Glioma \cite{b48,b49,b50,b51,b52} and Brats 2023 Meningioma \cite{b53} datasets are utilized for synthesizing MRI and brain tumor classification tasks. Brats 2023 Glioma and Meningioma challenges provide pre-gadolinium T1-weighted (T1), post-gadolinium T1-weighted (T1CE), T2-weighted (T2), T2-weighted fluid-attenuated inversion recovery (FLAIR) and segmented masks. The segmented tumors are labeled enhancing tumor, tumor core, and whole tumor. Axial samples of MRI slices including tumor and non-tumor regions are given in Figure 2. While the first MRI slices contain all tumor labels with maximum coverage, the slices in the 2\textsuperscript{nd} rows do not completely present whole tumor labels because of smaller coverage. Therefore, we select slices of MRI from each glioma and meningioma case with maximum tumor region coverage to learn the most relevant information for the pretraining and classification stages. On the other hand, 3\textsuperscript{rd} and the last row of Figure 2 show the healthy samples of MRI slices. While the presented samples of each slice do not contain any tumor region, the slices in the 3\textsuperscript{rd} row provide more contextual and relevant information about brain tissue. Therefore, we selected a healthy slice closer to the center slice of the brain MRI to capture more relevant information. Consequently, we arranged two sub-datasets for the pretraining stage and classification step. The top five slices with maximum tumor coverage have been selected from 1251 glioma cases and 1000 meningioma cases, along with the top five healthy samples based on their proximity to the center slice during the pretraining stage from a total of 2251 glioma and meningioma cases for the pretraining stage, called a BRaTS dataset (5x). In the classification stage, a singular MRI slice per case has been selected based on its depiction of the most relevant information for each tumor, and the dataset is called a basic BRaTS dataset. Additionally, to ensure a balanced training dataset during the classification stage, data augmentation is employed through MRI image synthesis using ResViT. This dataset is referred to as the augmented BRaTS dataset. Thus, the number of training datasets for glioma and meningioma classes is effectively doubled through the synthesis of images in the classification stage. This augmentation is accomplished by utilizing ResViT, pre-trained on the BRaTS dataset(5x) in the pretext task stage. On the other hand, Kaggle and Figshare brain tumor MRI datasets have also been used to demonstrate the learning process and effectiveness of the proposed model on different brain MRI datasets. The Figshare dataset contains 3064 T1-weighted images for glioma, meningioma, and pituitary cases, and the Kaggle dataset includes 7023 MRI images for no tumor, glioma, meningioma, and pituitary cases. The datasets have been split into a train set and a test set at a ratio of 80:20. The number of images for each dataset is presented in Table 2.

\begin{figure}[!ht]
\centerline{\includegraphics[width=0.75\columnwidth]{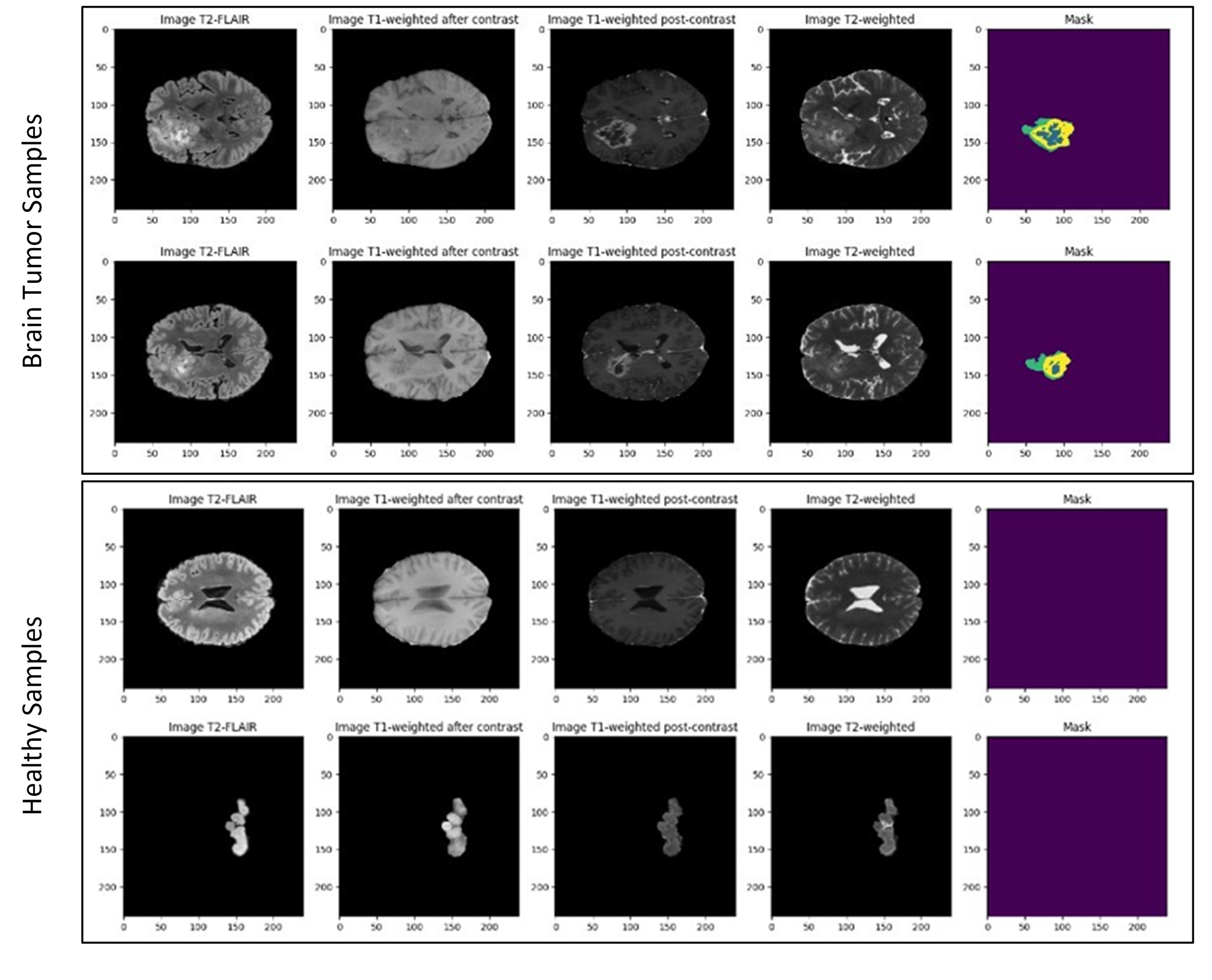}}
\caption{The several axial slices of MRI include labels and information of tumor region with coverage that all belong to the same case. The enhancing tumor (ET), the tumor core (TC), and the whole tumor (WT) are shown yellow, blue, and green, respectively.}
\label{fig2}

\end{figure}
\vspace{-0.4cm}
\begin{table*}[!ht]
\label{table2}
\scriptsize
\centering
\caption{The number of MRI images for each dataset, built with T1, T2, or Flair, used in the pretraining and classification stages.}
\begin{tabular}{|l|l|l|llll|}
\hline
&& \multicolumn{1}{c|}{\textbf{Pretraining stage}} & \multicolumn{4}{c|}{\textbf{Classification Stage}}                                                                                                   \\ \cline{1-7} 
                                  {\textbf{classes}} & {\textbf{subset}}                                   & \textbf{BRaTS   dataset(5x)}                     & {\textbf{Basic   BRaTS}} & {\textbf{Augmented   BRaTS}} & {\textbf{Kaggle}} & \textbf{Figshare} \\ \hline
No tumor                          & Train                            & 9005                                              & {1801}                   & {1801}                       & {1595}            & -                 \\ \hline
                                  & Test                             & 450                                               & {450}                    & {450}                        & {405}             & -                 \\ \hline
Glioma                            & Train                            & 5005                                              & {1001}                   & {2002}                       & {1321}            & 1,141             \\ \hline
                                  & Test                             & 250                                               & {250}                    & {200}                        & {300}             & 285               \\ \hline
Meningioma                        & Train                            & 4000                                              & {800}                    & {1600}                       & {1339}            & 566               \\ \hline
                                  & Test                             & 200                                               & {200}                    & {200}                        & {306}             & 142               \\ \hline
Pituitary                         & Train                            & -                                                 & {-}                      & {-}                          & {1457}            & 744               \\ \hline
                                  & Test                             & -                                                 & {-}                      & {-}                          & {300}             & 186               \\ \hline
\textbf{Total}                    & \textbf{Train}                   & \textbf{18,010}                                   & {\textbf{3602}}          & {\textbf{5403}}              & {\textbf{5712}}   & \textbf{2451}     \\ \hline
\textbf{}                         & \textbf{Test}                    & \textbf{900}                                      & {\textbf{900}}           & {\textbf{900}}               & {\textbf{1311}}   & \textbf{613}      \\ \hline
\end{tabular}
\end{table*}

\subsection{Experimental Setup}
The proposed deep learning model has been built on PyTorch Library. The deep models have been performed on A100-40GB GPU. While The BraTS dataset typically has a volume dimension of 240×240×155 voxels in NIfTI format, the Figshare dataset provides images with a resolution of 512×512 pixels in MATLAB (.mat) format. However, the Kaggle dataset provides 2D MRI images in various sizes. To standardize the resolution across datasets, the MRI image resolution is fixed at 256x256 pixels. The proposed and selected alternative deep learning models for MRI image synthesis have been implemented by the following settings: 256x256 image size, 1e-4 learning rate, Adam optimizer, and 100 epochs. The proposed and transfer learning models (mentioned in section 3. C) for classification have been set up with 256x256 image size, 16 batch sizes, Adam optimizer, 2e-5 learning rate, and 100 epochs. The pre-trained ResVİT model has been used for end-to-end fine-tuning and data augmentation in the classification stage. The proposed classifier model has been trained and tested separately on each sequence (T1, T2, and Flair). Therefore, pre-trained ResViT from T1 to T2, T2 to T1, and Flair to T1 models are transferred into the classifier model on T1, T2, and Flair, respectively. In addition, pre-trained ResViT from T1 to T2 model have been utilized for transfer learning of the proposed classifier model on Figshare and Kaggle Brain Tumor Datasets. The synthesis quality between source and synthetic MRI was calculated using peak signal-to-noise ratio (PSNR), structural similarity index \cite{b58}, and mean square error (MSE), of which the mean and standard deviations (std) were reported. The classification performance models were evaluated by accuracy, precision, recall, and F1 metrics, which were reported with weighted averages.

\subsection{Brain MRI Image Synthesis Results}

The MRI synthesis results are reported in Table 3. ResViT achieves PSNR of 25.391, SSIM of 0.878, MSE of 0.004 for T2 to T1 synthesis, PSNR of 25.663, SSIM of 0.884, MSE of 0.003 for T1 to T2 synthesis, and PSNR of 25.617, SSIM of 0.873, MSE of 0.004 for Flair to T1, and PSNR of 22.063, SSIM of 0.839, MSE of 0.008 for T1 to Flair. The ResViT demonstrates higher quality for synthesizing MRI images than other generative models (explained in Alternative Deep Learning Models for Comparative Study Section), especially the generator network constructed by Unet (A-Unet and pix2pix). The samples for MRI synthesis are given in Figure 3. The image-to-image translation models constructed by the ResNet, particularly ResViT, effectively captured and represented the tumor region with lower artifact and sharper tissue depiction than Unet-based models. 

On the other hand, ResViT and alternative image-to-image translation deep learning models are evaluated on various MRI sequences (T1, T2, and Flair). T1, T2, and Flair provide distinct information about brain tissue characteristics. While T1 highlights anatomical details such as white matter, gray matter, and cerebrospinal fluid, T2 provides differences in water content. Flair is a type of T2 sequence used to improve the visualization of lesions by suppressing cerebrospinal fluid (CSF) signals. The image-to-image translation models perform well across different T1 and T2 synthesis tasks. The quality of synthesis T1 and T2 images is slightly higher by approximately 3dB for PSNR than Flair synthesis. 

In summary, this pretraining stage is designed to enhance classification performance through fine-tuning and data augmentation with synthesized MRI images. For this purpose, ResViT is selected to adapt to the classification stage because ResViT outperforms and has a unique fusion of Residual CNN and transformer blocks to capture local and global features simultaneously. Moreover, ResViT synthesizes MRI images with lower artifacts, higher quality, and sharper brain tissue depiction, further supporting the brain classification task. 

\begin{table*}[!ht]
\centering
\label{table3}
\scriptsize
\caption{The results of generating synthetic images with generative deep models are evaluated by the mean and standard deviation (std) of signal-to-noise ratio (psnr), structural similarity index, and mean square error (mse) across the test images.}
\begin{tabular}{|l|lll|lll|lll|lll|}
\hline
\multirow{2}{*}{\textbf{model}} & \multicolumn{3}{c|}{\textbf{T2 -\textgreater T1}}                                                                                                                                                                                & \multicolumn{3}{c|}{\textbf{T1 -\textgreater T2}}                                                                                                                                                                                & \multicolumn{3}{c|}{\textbf{Flair-\textgreater{}T1}}                                                                                                                                                                             & \multicolumn{3}{c|}{\textbf{T1 -\textgreater Flair}}                                                                                                                                                                             \\ \cline{2-13} 
                                 & \multicolumn{1}{l|}{\textbf{\begin{tabular}[c]{@{}l@{}}PSNR\\ std\end{tabular}}} & \multicolumn{1}{l|}{\textbf{\begin{tabular}[c]{@{}l@{}}SSIM\\ std\end{tabular}}} & \textbf{\begin{tabular}[c]{@{}l@{}}MSE\\ std\end{tabular}} & \multicolumn{1}{l|}{\textbf{\begin{tabular}[c]{@{}l@{}}PSNR\\ std\end{tabular}}} & \multicolumn{1}{l|}{\textbf{\begin{tabular}[c]{@{}l@{}}SSIM\\ std\end{tabular}}} & \textbf{\begin{tabular}[c]{@{}l@{}}MSE\\ std\end{tabular}} & \multicolumn{1}{l|}{\textbf{\begin{tabular}[c]{@{}l@{}}PSNR\\ std\end{tabular}}} & \multicolumn{1}{l|}{\textbf{\begin{tabular}[c]{@{}l@{}}SSIM\\ std\end{tabular}}} & \textbf{\begin{tabular}[c]{@{}l@{}}MSE\\ std\end{tabular}} & \multicolumn{1}{l|}{\textbf{\begin{tabular}[c]{@{}l@{}}PSNR\\ std\end{tabular}}} & \multicolumn{1}{l|}{\textbf{\begin{tabular}[c]{@{}l@{}}SSIM\\ std\end{tabular}}} & \textbf{\begin{tabular}[c]{@{}l@{}}MSE\\ std\end{tabular}} \\ \hline
\multirow{2}{*}{A-Unet}          & \multicolumn{1}{l|}{24.383}                                                      & \multicolumn{1}{l|}{0.856}                                                       & 0.005                                                      & \multicolumn{1}{l|}{23.968}                                                      & \multicolumn{1}{l|}{0.852}                                                       & 0.005                                                      & \multicolumn{1}{l|}{23.574}                                                      & \multicolumn{1}{l|}{0.832}                                                       & 0.006                                                      & \multicolumn{1}{l|}{20.830}                                                      & \multicolumn{1}{l|}{0.803}                                                       & 0.010                                                      \\ \cline{2-13} 
                                 & \multicolumn{1}{l|}{±3.149}                                                      & \multicolumn{1}{l|}{±0.042}                                                      & ±0.005                                                     & \multicolumn{1}{l|}{±2.046}                                                      & \multicolumn{1}{l|}{±0.057}                                                      & ±0.003                                                     & \multicolumn{1}{l|}{±3.204}                                                      & \multicolumn{1}{l|}{±0.052}                                                      & ±0.006                                                     & \multicolumn{1}{l|}{±2.812}                                                      & \multicolumn{1}{l|}{±0.062}                                                      & ±0.007                                                     \\ \hline
\multirow{2}{*}{Resnet-9}        & \multicolumn{1}{l|}{25.089}                                                      & \multicolumn{1}{l|}{0.870}                                                       & 0.004                                                      & \multicolumn{1}{l|}{25.105}                                                      & \multicolumn{1}{l|}{0.875}                                                       & 0.004                                                      & \multicolumn{1}{l|}{24.961}                                                      & \multicolumn{1}{l|}{0.867}                                                       & 0.004                                                      & \multicolumn{1}{l|}{\textbf{22.145}}                                             & \multicolumn{1}{l|}{0.838}                                                       & \textbf{0.008}                                             \\ \cline{2-13} 
                                 & \multicolumn{1}{l|}{±3.194}                                                      & \multicolumn{1}{l|}{±0.039}                                                      & ±0.005                                                     & \multicolumn{1}{l|}{±2.150}                                                      & \multicolumn{1}{l|}{±0.052}                                                      & ±0.002                                                     & \multicolumn{1}{l|}{±3.158}                                                      & \multicolumn{1}{l|}{±0.039}                                                      & ±0.005                                                     & \multicolumn{1}{l|}{±2.848}                                                      & \multicolumn{1}{l|}{±0.051}                                                      & ±0.006                                                     \\ \hline
\multirow{2}{*}{pix2pix}         & \multicolumn{1}{l|}{24.023}                                                      & \multicolumn{1}{l|}{0.821}                                                       & 0.005                                                      & \multicolumn{1}{l|}{24.619}                                                      & \multicolumn{1}{l|}{0.860}                                                       & 0.004                                                      & \multicolumn{1}{l|}{24.945}                                                      & \multicolumn{1}{l|}{0.854}                                                       & 0.005                                                      & \multicolumn{1}{l|}{21.469}                                                      & \multicolumn{1}{l|}{0.831}                                                       & 0.010                                                      \\ \cline{2-13} 
                                 & \multicolumn{1}{l|}{±3.087}                                                      & \multicolumn{1}{l|}{±0.045}                                                      & ±0.006                                                     & \multicolumn{1}{l|}{±2.113}                                                      & \multicolumn{1}{l|}{±0.052}                                                      & ±0.003                                                     & \multicolumn{1}{l|}{±3.269}                                                      & \multicolumn{1}{l|}{±0.041}                                                      & ±0.006                                                     & \multicolumn{1}{l|}{±3.088}                                                      & \multicolumn{1}{l|}{±0.071}                                                      & ±0.010                                                     \\ \hline
\multirow{2}{*}{pGAN}            & \multicolumn{1}{l|}{24.943}                                                      & \multicolumn{1}{l|}{0.870}                                                       & 0.004                                                      & \multicolumn{1}{l|}{24.208}                                                      & \multicolumn{1}{l|}{0.868}                                                       & 0.004                                                      & \multicolumn{1}{l|}{24.327}                                                      & \multicolumn{1}{l|}{0.863}                                                       & 0.005                                                      & \multicolumn{1}{l|}{20.355}                                                      & \multicolumn{1}{l|}{0.835}                                                       & 0.009                                                      \\ \cline{2-13} 
                                 & \multicolumn{1}{l|}{±3.212}                                                      & \multicolumn{1}{l|}{±0.039}                                                      & ±0.005                                                     & \multicolumn{1}{l|}{±2.193}                                                      & \multicolumn{1}{l|}{±0.053}                                                      & ±0.003                                                     & \multicolumn{1}{l|}{±3.256}                                                      & \multicolumn{1}{l|}{±0.041}                                                      & ±0.005                                                     & \multicolumn{1}{l|}{±3.210}                                                      & \multicolumn{1}{l|}{±0.053}                                                      & ±0.009                                                     \\ \hline
\multirow{2}{*}{\textbf{ResViT}} & \multicolumn{1}{l|}{\textbf{25.391}}                                             & \multicolumn{1}{l|}{\textbf{0.878}}                                              & \textbf{0.004}                                             & \multicolumn{1}{l|}{\textbf{25.663}}                                             & \multicolumn{1}{l|}{\textbf{0.884}}                                              & \textbf{0.003}                                             & \multicolumn{1}{l|}{\textbf{25.617}}                                             & \multicolumn{1}{l|}{\textbf{0.873}}                                              & \textbf{0.004}                                             & \multicolumn{1}{l|}{22.063}                                                      & \multicolumn{1}{l|}{\textbf{0.839}}                                              & \textbf{0.008}                                             \\ \cline{2-13} 
                                 & \multicolumn{1}{l|}{±3.289}                                                      & \multicolumn{1}{l|}{±0.039}                                                      & ±0.005                                                     & \multicolumn{1}{l|}{±2.237}                                                      & \multicolumn{1}{l|}{±0.052}                                                      & ±0.002                                                     & \multicolumn{1}{l|}{±3.268}                                                      & \multicolumn{1}{l|}{±0.040}                                                      & ±0.005                                                     & \multicolumn{1}{l|}{±2.945}                                                      & \multicolumn{1}{l|}{±0.051}                                                      & ±0.008                                                     \\ \hline
\end{tabular}
\end{table*}

\begin{figure}[!ht]
\centerline{\includegraphics[width=\columnwidth]{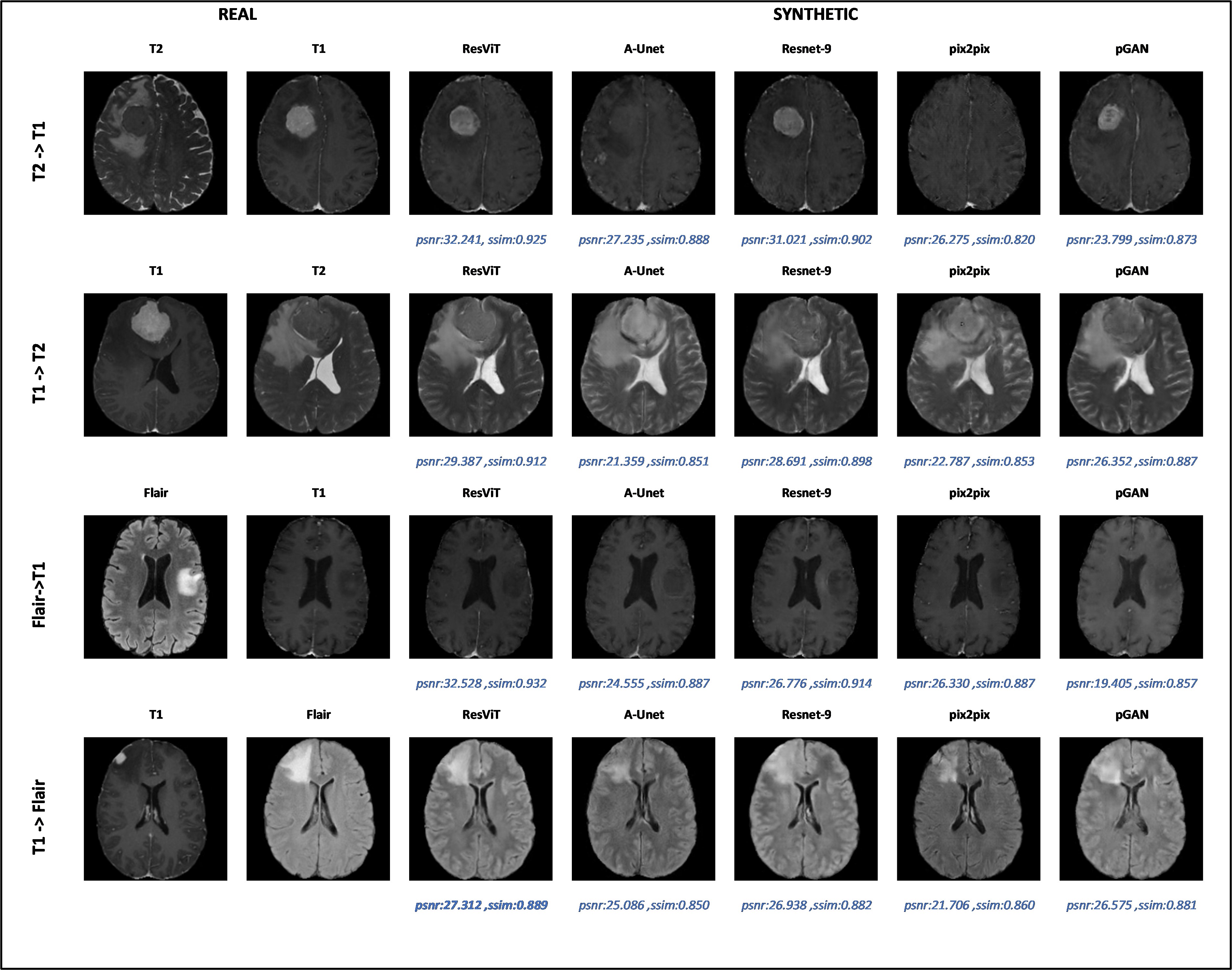}}
\caption{The samples of synthesis MRI from T2 to T1, T2 to T1, Flair to T1, and T1 to Flair.}
\label{fig3}
\end{figure}

\subsection{Classification Results}

The proposed ResViT-based classifier and transfer learning models have been performed and evaluated on the basic BraTS, augmented BraTS, Figshare, and Kaggle datasets. Furthermore, the classification models train and test separately for each MRI sequence (T1, T2, and Flair). The results for each MRI sequence on the basic BraTS dataset have been reported in Table 4. ResViT model demonstrates superior performance compared to ConvNeXtTiny, Resnet101, Densenet121, Residual CNN, and ViT models for each Brain MRI sequence. The proposed ResViT-based classifier model achieves an accuracy of 88.89\% for T1, 86.22\% for Flair, and 83.44\% for T2 on the basic Brats dataset. The results clearly indicate that the generative self-supervised learning strategy significantly improved classification performance. In terms of fine-tuning models, transfer learning of the pre-trained ResViT model on the BraTS MRI dataset into the proposed classifier model provides higher performance for each dataset and sequence than other pre-trained models on ImageNet. Although Densenet121 obtains the best scores among the state-of-the-art transfer learning models pre-trained on ImageNet (ConvNeXtTiny, Resnet101, Densenet121), the proposed model has slightly higher accuracy by approximately 9 points for T1, 7 points for Flair, 8 points for T2. Furthermore, the proposed ResViT-based generative self-supervised learning achieves better success than the pre-trained base (R50+ViT-B\_16, ViT-B\_16) and large (ViT-L\_16) transformer models on ImageNet. 
\vspace{-0.1cm}
\begin{table*}[!ht]
\scriptsize
\centering
\label{table4}
\caption{The classification results of deep learning models for T1, T2, and Flair sequence on the basic Brats dataset.}

\begin{tabular}{|l|l|llll|llll|llll|}
\hline
\multirow{2}{*}{\textbf{Model}}                                & \multirow{2}{*}{\textbf{pretraining}} & \multicolumn{4}{c|}{\textbf{T1}}                                                                                                      & \multicolumn{4}{c|}{\textbf{FLAIR}}                                                                                                   & \multicolumn{4}{c|}{\textbf{T2}}                                                                                                      \\ \cline{3-14} 
                                                               &                                       & \multicolumn{1}{l|}{\textbf{acc}}   & \multicolumn{1}{l|}{\textbf{precision}} & \multicolumn{1}{l|}{\textbf{recall}} & \textbf{F1}    & \multicolumn{1}{l|}{\textbf{acc}}   & \multicolumn{1}{l|}{\textbf{precision}} & \multicolumn{1}{l|}{\textbf{recall}} & \textbf{F1}    & \multicolumn{1}{l|}{\textbf{acc}}   & \multicolumn{1}{l|}{\textbf{precision}} & \multicolumn{1}{l|}{\textbf{recall}} & \textbf{F1}    \\  \hline
ConvNeXtTiny                                          & ImageNet                     & \multicolumn{1}{l|}{59.67} & \multicolumn{1}{l|}{61.28}     & \multicolumn{1}{l|}{59.56}  & 54.83 & \multicolumn{1}{l|}{60.67} & \multicolumn{1}{l|}{59.22}     & \multicolumn{1}{l|}{60.22}  & 59.22 & \multicolumn{1}{l|}{55.56} & \multicolumn{1}{l|}{57.83}     & \multicolumn{1}{l|}{55.56}  & 56.22 \\ \hline
Resnet101                                             & ImageNet                     & \multicolumn{1}{l|}{64.44} & \multicolumn{1}{l|}{67.83}     & \multicolumn{1}{l|}{64.61}  & 63.33 & \multicolumn{1}{l|}{71.67} & \multicolumn{1}{l|}{71.83}     & \multicolumn{1}{l|}{71.67}  & 70.39 & \multicolumn{1}{l|}{62.78} & \multicolumn{1}{l|}{62.39}     & \multicolumn{1}{l|}{63.11}  & 61.83 \\ \hline
Densenet121                                           & ImageNet                     & \multicolumn{1}{l|}{79.78} & \multicolumn{1}{l|}{80.50}     & \multicolumn{1}{l|}{79.89}  & 79.33 & \multicolumn{1}{l|}{79.00} & \multicolumn{1}{l|}{79.78}     & \multicolumn{1}{l|}{78.94}  & 79.33 & \multicolumn{1}{l|}{75.33} & \multicolumn{1}{l|}{75.33}     & \multicolumn{1}{l|}{75.33}  & 75.17 \\ \hline
Residual CNN                                          & w/o fine-tuning              & \multicolumn{1}{l|}{80.56} & \multicolumn{1}{l|}{80.33}     & \multicolumn{1}{l|}{80.89}  & 80.11 & \multicolumn{1}{l|}{77.11} & \multicolumn{1}{l|}{76.56}     & \multicolumn{1}{l|}{77.00}  & 75.83 & \multicolumn{1}{l|}{69.22} & \multicolumn{1}{l|}{68.17}     & \multicolumn{1}{l|}{69.00}  & 68.22 \\ \hline
\multirow{4}{*}{ViT}                                  & w/o finetuning               & \multicolumn{1}{l|}{79.44} & \multicolumn{1}{l|}{79.17}     & \multicolumn{1}{l|}{79.50}  & 79.33 & \multicolumn{1}{l|}{79.00} & \multicolumn{1}{l|}{78.33}     & \multicolumn{1}{l|}{79.17}  & 78.28 & \multicolumn{1}{l|}{71.67} & \multicolumn{1}{l|}{70.50}     & \multicolumn{1}{l|}{71.67}  & 70.83 \\ \cline{2-14} 
                                                      & R50+ViT-B\_16                & \multicolumn{1}{l|}{83.22} & \multicolumn{1}{l|}{83.78}     & \multicolumn{1}{l|}{83.06}  & 83.06 & \multicolumn{1}{l|}{78.33} & \multicolumn{1}{l|}{77.61}     & \multicolumn{1}{l|}{78.33}  & 77.61 & \multicolumn{1}{l|}{74.44} & \multicolumn{1}{l|}{75.17}     & \multicolumn{1}{l|}{74.61}  & 74.39 \\ \cline{2-14} 
                                                      & ViT-B\_16                    & \multicolumn{1}{l|}{79.56} & \multicolumn{1}{l|}{79.78}     & \multicolumn{1}{l|}{79.33}  & 79.44 & \multicolumn{1}{l|}{80.78} & \multicolumn{1}{l|}{80.28}     & \multicolumn{1}{l|}{80.83}  & 80.11 & \multicolumn{1}{l|}{69.56} & \multicolumn{1}{l|}{69.00}     & \multicolumn{1}{l|}{69.61}  & 69.17 \\ \cline{2-14} 
                                                      & ViT-L\_16                    & \multicolumn{1}{l|}{81.89} & \multicolumn{1}{l|}{82.00}     & \multicolumn{1}{l|}{82.17}  & 81.94 & \multicolumn{1}{l|}{79.44} & \multicolumn{1}{l|}{78.78}     & \multicolumn{1}{l|}{79.39}  & 78.17 & \multicolumn{1}{l|}{76.56} & \multicolumn{1}{l|}{75.33}     & \multicolumn{1}{l|}{76.94}  & 75.56 \\ \hline
\multirow{5}{*}{\textbf{\begin{tabular}[c]{@{}l@{}}The proposed \\ ResViT-based\\ classifier\end{tabular}}} & w/o fine-tuning              & \multicolumn{1}{l|}{81.44} & \multicolumn{1}{l|}{80.89}     & \multicolumn{1}{l|}{81.28}  & 80.83 & \multicolumn{1}{l|}{77.56} & \multicolumn{1}{l|}{76.83}     & \multicolumn{1}{l|}{77.44}  & 76.67 & \multicolumn{1}{l|}{75.67} & \multicolumn{1}{l|}{74.67}     & \multicolumn{1}{l|}{75.56}  & 74.39 \\ \cline{2-14} 
                                                      & R50+ViT-B\_16                & \multicolumn{1}{l|}{83.56} & \multicolumn{1}{l|}{83.67}     & \multicolumn{1}{l|}{83.83}  & 83.39 & \multicolumn{1}{l|}{78.78} & \multicolumn{1}{l|}{78.50}     & \multicolumn{1}{l|}{78.94}  & 78.61 & \multicolumn{1}{l|}{75.67} & \multicolumn{1}{l|}{75.94}     & \multicolumn{1}{l|}{77.11}  & 76.17 \\ \cline{2-14} 
                                                      & ViT-B\_16                    & \multicolumn{1}{l|}{84.11} & \multicolumn{1}{l|}{84.50}     & \multicolumn{1}{l|}{84.00}  & 83.72 & \multicolumn{1}{l|}{78.56} & \multicolumn{1}{l|}{78.83}     & \multicolumn{1}{l|}{78.61}  & 78.22 & \multicolumn{1}{l|}{76.67} & \multicolumn{1}{l|}{75.72}     & \multicolumn{1}{l|}{76.67}  & 76.33 \\ \cline{2-14} 
                                                      & ViT-L\_16                    & \multicolumn{1}{l|}{85.56} & \multicolumn{1}{l|}{85.06}     & \multicolumn{1}{l|}{85.89}  & 85.33 & \multicolumn{1}{l|}{78.33} & \multicolumn{1}{l|}{77.33}     & \multicolumn{1}{l|}{78.72}  & 77.44 & \multicolumn{1}{l|}{76.78} & \multicolumn{1}{l|}{76.50}     & \multicolumn{1}{l|}{76.94}  & 76.22 \\ \cline{2-14} 
                                                      & \textbf{ResViT-Proposed}     & \multicolumn{1}{l|}{\textbf{88.89}} & \multicolumn{1}{l|}{\textbf{89.22}} & \multicolumn{1}{l|}{\textbf{88.67}} & \textbf{88.83} & \multicolumn{1}{l|}{\textbf{86.22}} & \multicolumn{1}{l|}{\textbf{85.78}} & \multicolumn{1}{l|}{\textbf{86.22}} & \textbf{86.17} & \multicolumn{1}{l|}{\textbf{83.44}} & \multicolumn{1}{l|}{\textbf{82.89}} & \multicolumn{1}{l|}{\textbf{83.33}} & \textbf{83.00} \\ \hline
\end{tabular}

\end{table*}

\begin{table*}[!ht]
\scriptsize
\label{table5}

\centering
\caption{The classification results of deep learning models for T1, T2, and Flair on the augmented Brats dataset.}
\begin{tabular}{|l|l|llll|llll|llll|}
\hline
\multirow{2}{*}{\textbf{Model}}                                & \multirow{2}{*}{\textbf{pretraining}} & \multicolumn{4}{c|}{\textbf{T1}}                                                                                                      & \multicolumn{4}{c|}{\textbf{FLAIR}}                                                                                                   & \multicolumn{4}{c|}{\textbf{T2}}                                                                                                      \\ \cline{3-14} 
                                                               &                                       & \multicolumn{1}{l|}{\textbf{acc}}   & \multicolumn{1}{l|}{\textbf{precision}} & \multicolumn{1}{l|}{\textbf{recall}} & \textbf{F1}    & \multicolumn{1}{l|}{\textbf{acc}}   & \multicolumn{1}{l|}{\textbf{precision}} & \multicolumn{1}{l|}{\textbf{recall}} & \textbf{F1}    & \multicolumn{1}{l|}{\textbf{acc}}   & \multicolumn{1}{l|}{\textbf{precision}} & \multicolumn{1}{l|}{\textbf{recall}} & \textbf{F1}    \\ \hline
ConvNeXtTiny                                                   & ImageNet                              & \multicolumn{1}{l|}{36.56}          & \multicolumn{1}{l|}{53.39}              & \multicolumn{1}{l|}{36.72}           & 28.89          & \multicolumn{1}{l|}{44.11}          & \multicolumn{1}{l|}{54.44}              & \multicolumn{1}{l|}{43.94}           & 42.11          & \multicolumn{1}{l|}{55.56}          & \multicolumn{1}{l|}{57.83}              & \multicolumn{1}{l|}{55.56}           & 56.22          \\ \hline
Resnet101                                                      & ImageNet                              & \multicolumn{1}{l|}{60.67}          & \multicolumn{1}{l|}{66.94}              & \multicolumn{1}{l|}{60.56}           & 61.11          & \multicolumn{1}{l|}{71.78}          & \multicolumn{1}{l|}{73.72}              & \multicolumn{1}{l|}{71.67}           & 72.22          & \multicolumn{1}{l|}{62.89}          & \multicolumn{1}{l|}{63.44}              & \multicolumn{1}{l|}{63.06}           & 62.33          \\ \hline
Densenet121                                                    & ImageNet                              & \multicolumn{1}{l|}{78.78}          & \multicolumn{1}{l|}{78.89}              & \multicolumn{1}{l|}{78.94}           & 79.06          & \multicolumn{1}{l|}{74.44}          & \multicolumn{1}{l|}{75.83}              & \multicolumn{1}{l|}{74.44}           & 74.89          & \multicolumn{1}{l|}{72.67}          & \multicolumn{1}{l|}{74.22}              & \multicolumn{1}{l|}{72.50}           & 72.94          \\ \hline
Residual CNN                                                   & w/o fine-tuning                       & \multicolumn{1}{l|}{80.22}          & \multicolumn{1}{l|}{80.17}              & \multicolumn{1}{l|}{80.33}           & 80.39          & \multicolumn{1}{l|}{76.11}          & \multicolumn{1}{l|}{74.94}              & \multicolumn{1}{l|}{75.83}           & 75.06          & \multicolumn{1}{l|}{76.89}          & \multicolumn{1}{l|}{76.89}              & \multicolumn{1}{l|}{76.56}           & 76.72          \\ \hline
\multirow{4}{*}{ViT}                                           & w/o finetuning                        & \multicolumn{1}{l|}{76.56}          & \multicolumn{1}{l|}{77.50}              & \multicolumn{1}{l|}{76.72}           & 76.61          & \multicolumn{1}{l|}{77.44}          & \multicolumn{1}{l|}{77.39}              & \multicolumn{1}{l|}{77.50}           & 77.44          & \multicolumn{1}{l|}{77.44}          & \multicolumn{1}{l|}{76.89}              & \multicolumn{1}{l|}{77.72}           & 77.06          \\ \cline{2-14} 
                                                               & R50+ViT-B\_16                         & \multicolumn{1}{l|}{80.00}          & \multicolumn{1}{l|}{79.28}              & \multicolumn{1}{l|}{80.00}           & 79.50          & \multicolumn{1}{l|}{80.22}          & \multicolumn{1}{l|}{79.61}              & \multicolumn{1}{l|}{80.33}           & 79.50          & \multicolumn{1}{l|}{78.11}          & \multicolumn{1}{l|}{77.78}              & \multicolumn{1}{l|}{78.39}           & 77.83          \\ \cline{2-14} 
                                                               & ViT-B\_16                             & \multicolumn{1}{l|}{80.11}          & \multicolumn{1}{l|}{79.11}              & \multicolumn{1}{l|}{80.11}           & 79.39          & \multicolumn{1}{l|}{79.67}          & \multicolumn{1}{l|}{79.00}              & \multicolumn{1}{l|}{79.44}           & 79.28          & \multicolumn{1}{l|}{77.78}          & \multicolumn{1}{l|}{79.72}              & \multicolumn{1}{l|}{80.17}           & 80.44          \\ \cline{2-14} 
                                                               & ViT-L\_16                             & \multicolumn{1}{l|}{85.56}          & \multicolumn{1}{l|}{80.17}              & \multicolumn{1}{l|}{80.61}           & 80.00          & \multicolumn{1}{l|}{79.56}          & \multicolumn{1}{l|}{79.06}              & \multicolumn{1}{l|}{79.44}           & 79.11          & \multicolumn{1}{l|}{79.11}          & \multicolumn{1}{l|}{78.94}              & \multicolumn{1}{l|}{78.94}           & 78.83          \\ \hline
\multirow{5}{*}{\textbf{\begin{tabular}[c]{@{}l@{}}The proposed \\ ResViT-based\\ classifier\end{tabular}}}& w/o fine-tuning                       & \multicolumn{1}{l|}{81.67}          & \multicolumn{1}{l|}{82.83}              & \multicolumn{1}{l|}{81.56}           & 81.83          & \multicolumn{1}{l|}{80.00}          & \multicolumn{1}{l|}{79.28}              & \multicolumn{1}{l|}{79.83}           & 79.44          & \multicolumn{1}{l|}{80.89}          & \multicolumn{1}{l|}{81.44}              & \multicolumn{1}{l|}{80.78}           & 81.22          \\ \cline{2-14} 
                                                               & R50+ViT-B\_16                         & \multicolumn{1}{l|}{83.11}          & \multicolumn{1}{l|}{82.83}              & \multicolumn{1}{l|}{83.22}           & 82.67          & \multicolumn{1}{l|}{78.44}          & \multicolumn{1}{l|}{77.78}              & \multicolumn{1}{l|}{78.39}           & 77.83          & \multicolumn{1}{l|}{79.22}          & \multicolumn{1}{l|}{79.28}              & \multicolumn{1}{l|}{79.39}           & 78.83          \\ \cline{2-14} 
                                                               & ViT-B\_16                             & \multicolumn{1}{l|}{86.44}          & \multicolumn{1}{l|}{86.61}              & \multicolumn{1}{l|}{86.67}           & 86.39          & \multicolumn{1}{l|}{80.78}          & \multicolumn{1}{l|}{80.72}              & \multicolumn{1}{l|}{80.83}           & 80.28          & \multicolumn{1}{l|}{79.44}          & \multicolumn{1}{l|}{79.39}              & \multicolumn{1}{l|}{79.61}           & 79.22          \\ \cline{2-14} 
                                                               & ViT-L\_16                             & \multicolumn{1}{l|}{83.44}          & \multicolumn{1}{l|}{83.39}              & \multicolumn{1}{l|}{83.39}           & 83.28          & \multicolumn{1}{l|}{80.33}          & \multicolumn{1}{l|}{80.44}              & \multicolumn{1}{l|}{79.67}           & 79.56          & \multicolumn{1}{l|}{79.67}          & \multicolumn{1}{l|}{79.17}              & \multicolumn{1}{l|}{79.56}           & 78.89          \\ \cline{2-14} 
                                                               & \textbf{ResViT-Proposed}              & \multicolumn{1}{l|}{\textbf{90.56}} & \multicolumn{1}{l|}{\textbf{91.06}}     & \multicolumn{1}{l|}{\textbf{90.72}}  & \textbf{90.78} & \multicolumn{1}{l|}{\textbf{88.44}} & \multicolumn{1}{l|}{\textbf{88.61}}     & \multicolumn{1}{l|}{\textbf{88.28}}  & \textbf{88.33} & \multicolumn{1}{l|}{\textbf{88.89}} & \multicolumn{1}{l|}{\textbf{89.17}}     & \multicolumn{1}{l|}{\textbf{88.94}}  & \textbf{89.06} \\ \hline
\end{tabular}
\end{table*}

On the other hand, the synthetic MRI images by ResViT have been utilized as data augmentation in the classification stage. Therefore, the augmented BraTS dataset includes synthesized MRI images for glioma and meningioma cases. Table 5 presents the results of classification models on the augmented Brats dataset. The results demonstrate that using synthetic images by ResViT enhances the classification performance of the proposed model. The accuracy of the proposed model increased from 88.89\% to 90.56\% for T1, from 86.22\% to 88.44\% for Flair, and from 83.44\% to 88.89\% for T2. Furthermore, combining Residual CNN and ViT obtains more effective performance than their separate usage (Residual CNN, ViT). Moreover, the proposed model exhibits the highest achievement on the T1 sequence among various sequence types (T1, Flair, T2) with an accuracy of 88.89\% for the basic Brats dataset and 90.56\% for the augmented Brats dataset. Consequently, utilizing a self-learning strategy and synthesized MRI together images distinctly enhances the classification performance of the proposed ResViT-based classifier model.

The classification models have been performed on Figshare and Kaggle Brain Tumor Dataset to assess the effectiveness of the proposed model on different brain MRI datasets. Table 6 shows the results of classification models on Figshare and Kaggle Brain Tumor Dataset. The pre-trained ResViT from T1 to T2 model has been selected owing to the highest performance on T1 for BraTs dataset, then transferred into the classifier model on Figshare and Kaggle datasets. The ResViT-based classifier model exhibits remarkable performance, surpassing other pre-trained models on natural datasets, with an accuracy of 98.53\% on the Figshare and 98.47\% on the Kaggle brain tumor datasets. These results demonstrate that leveraging a pre-trained model on the BraTs dataset rather than the natural dataset is an effective approach for transfer learning on various MRI datasets, indicating the versatility and generalizability of the ResViT-based classifier. The brain tumor classification studies commonly evaluate on Figshare dataset. Table 7 compares the proposed model with previous studies for Figshare dataset. The proposed model achieves a higher accuracy score of 98.53 on Figshare dataset against previous studies that are Swati \emph{et al.} \cite{b2}, Deepak \emph{et al.}\cite{b12}, Alshayeji \emph{et al.} \cite{b16}, Kakarla \emph{et al.} \cite{b17}, Kumar \emph{et al.}\cite{b18} and Ferdous \emph{et al.} \cite{b27}.

\begin{table*}[!ht]
\scriptsize
\label{table6}

\centering
\caption{The classification results of deep learning models on Figshare and Kaggle Brain Tumor Dataset.}
\begin{tabular}{|l|l|llll|llll|}
\hline
\multirow{2}{*}{\textbf{Model}}                                                                           & \multirow{2}{*}{\textbf{pretraining}} & \multicolumn{4}{c|}{\textbf{Figshare dataset}}                                                                                        & \multicolumn{4}{c|}{\textbf{Kaggle dataset}}                                                                                          \\ \cline{3-10} 
                                                                                                          &                                       & \multicolumn{1}{l|}{\textbf{acc}}   & \multicolumn{1}{l|}{\textbf{precision}} & \multicolumn{1}{l|}{\textbf{recall}} & \textbf{F1}    & \multicolumn{1}{l|}{\textbf{acc}}   & \multicolumn{1}{l|}{\textbf{precision}} & \multicolumn{1}{l|}{\textbf{recall}} & \textbf{F1}    \\ \hline
ConvNeXtTiny                                                                                              & ImageNet                              & \multicolumn{1}{l|}{81.89}          & \multicolumn{1}{l|}{81.52}              & \multicolumn{1}{l|}{82.08}           & 81.19          & \multicolumn{1}{l|}{89.93}          & \multicolumn{1}{l|}{89.83}              & \multicolumn{1}{l|}{89.95}           & 89.93          \\ \hline
Resnet101                                                                                                 & ImageNet                              & \multicolumn{1}{l|}{84.67}          & \multicolumn{1}{l|}{85.89}              & \multicolumn{1}{l|}{84.46}           & 84.86          & \multicolumn{1}{l|}{87.87}          & \multicolumn{1}{l|}{89.63}              & \multicolumn{1}{l|}{87.84}           & 87.31          \\ \hline
Densenet121                                                                                               & ImageNet                              & \multicolumn{1}{l|}{90.54}          & \multicolumn{1}{l|}{90.61}              & \multicolumn{1}{l|}{90.65}           & 90.13          & \multicolumn{1}{l|}{96.72}          & \multicolumn{1}{l|}{96.98}              & \multicolumn{1}{l|}{96.78}           & 96.77          \\ \hline
Residual CNN                                                                                              & w/o fine-tuning                       & \multicolumn{1}{l|}{95.43}          & \multicolumn{1}{l|}{95.70}              & \multicolumn{1}{l|}{95.52}           & 95.38          & \multicolumn{1}{l|}{94.66}          & \multicolumn{1}{l|}{94.53}              & \multicolumn{1}{l|}{94.45}           & 94.61          \\ \hline
\multirow{4}{*}{ViT}                                                                                      & w/o finetuning                        & \multicolumn{1}{l|}{93.31}          & \multicolumn{1}{l|}{93.08}              & \multicolumn{1}{l|}{93.27}           & 93.29          & \multicolumn{1}{l|}{94.81}          & \multicolumn{1}{l|}{94.84}              & \multicolumn{1}{l|}{94.92}           & 94.84          \\ \cline{2-10} 
                                                                                                          & R50+ViT-B\_16                         & \multicolumn{1}{l|}{95.27}          & \multicolumn{1}{l|}{95.47}              & \multicolumn{1}{l|}{94.83}           & 95.38          & \multicolumn{1}{l|}{96.80}          & \multicolumn{1}{l|}{96.91}              & \multicolumn{1}{l|}{96.54}           & 96.99          \\ \cline{2-10} 
                                                                                                          & ViT-B\_16                             & \multicolumn{1}{l|}{91.84}          & \multicolumn{1}{l|}{92.76}              & \multicolumn{1}{l|}{91.80}           & 91.97          & \multicolumn{1}{l|}{94.89}          & \multicolumn{1}{l|}{95.07}              & \multicolumn{1}{l|}{94.92}           & 94.84          \\ \cline{2-10} 
                                                                                                          & ViT-L\_16                             & \multicolumn{1}{l|}{93.47}          & \multicolumn{1}{l|}{93.31}              & \multicolumn{1}{l|}{93.43}           & 93.22          & \multicolumn{1}{l|}{93.59}          & \multicolumn{1}{l|}{93.67}              & \multicolumn{1}{l|}{93.55}           & 93.53          \\ \hline
\multirow{5}{*}{\textbf{\begin{tabular}[c]{@{}l@{}}The proposed \\ ResViT-based classifier\end{tabular}}} & w/o fine-tuning                       & \multicolumn{1}{l|}{94.94}          & \multicolumn{1}{l|}{95.16}              & \multicolumn{1}{l|}{95.13}           & 95.15          & \multicolumn{1}{l|}{95.12}          & \multicolumn{1}{l|}{95.14}              & \multicolumn{1}{l|}{95.38}           & 94.99          \\ \cline{2-10} 
                                                                                                          & R50+ViT-B\_16                         & \multicolumn{1}{l|}{95.11}          & \multicolumn{1}{l|}{94.93}              & \multicolumn{1}{l|}{95.13}           & 95.15          & \multicolumn{1}{l|}{94.89}          & \multicolumn{1}{l|}{94.84}              & \multicolumn{1}{l|}{94.90}           & 94.98          \\ \cline{2-10} 
                                                                                                          & ViT-B\_16                             & \multicolumn{1}{l|}{92.99}          & \multicolumn{1}{l|}{92.70}              & \multicolumn{1}{l|}{92.97}           & 92.99          & \multicolumn{1}{l|}{96.11}          & \multicolumn{1}{l|}{96.15}              & \multicolumn{1}{l|}{96.30}           & 95.99          \\ \cline{2-10} 
                                                                                                          & ViT-L\_16                             & \multicolumn{1}{l|}{94.94}          & \multicolumn{1}{l|}{94.63}              & \multicolumn{1}{l|}{95.13}           & 95.15          & \multicolumn{1}{l|}{94.20}          & \multicolumn{1}{l|}{94.59}              & \multicolumn{1}{l|}{93.93}           & 94.15          \\ \cline{2-10} 
                                                                                                          & \textbf{ResViT-Proposed}              & \multicolumn{1}{l|}{\textbf{98.53}} & \multicolumn{1}{l|}{\textbf{98.54}}     & \multicolumn{1}{l|}{\textbf{98.54}}  & \textbf{98.54} & \multicolumn{1}{l|}{\textbf{98.47}} & \multicolumn{1}{l|}{\textbf{98.45}}     & \multicolumn{1}{l|}{\textbf{98.61}}  & \textbf{98.53} \\ \hline
\end{tabular}
\end{table*}

\begin{table*}[!ht]
\scriptsize
\label{table7}

\centering
\caption{Comparative study in terms of the accuracy with the previous studies on Figshare dataset.}
\begin{tabular}{|l|l|l|l|}
\hline
\textbf{references}         & \textbf{year} & \textbf{method}                                                                                                                                  & \textbf{accuracy} \\ \hline
Swati \emph{et al.} {}\cite{b2}{}        & 2019          & pre-trained VGG19                                                                                                                                & 94.82             \\ \hline
Deepak \emph{et al.} {}\cite{b12}{}      & 2019          & transfer learning with GoogLeNet                                                                                                                 & 97.10             \\ \hline
Alshayeji \emph{et al.} {}\cite{b16}{}   & 2021          & CNN by using optimization for   hyperparameters                                                                                                  & 97.37             \\ \hline
Kakarla \emph{et al.} {}\cite{b17}{}      & 2021          & CNN and average pooling                                                                                                                          & 97.42             \\ \hline
Kumar \emph{et al.} {}\cite{b18}{}        & 2021          & ResNet‑50 and   average pooling                                                                                                                  & 97.48             \\ \hline
Ferdous \emph{et al.} {}\cite{b27}{}    & 2023          & LCDEiT by combining transformer and CNN                                                                                                          & 98.11             \\ \hline
\textbf{The proposed model} & \textbf{}     & \textbf{\begin{tabular}[c]{@{}l@{}}a Residual Vision Transformer-based generative SSL model by combining transformer and ResNet\end{tabular}} & \textbf{98.53}    \\ \hline
\end{tabular}
\end{table*}

\section{Conclusion}
This study introduced a self-supervised learning model consisting of a pre-trained ResViT model for MRI synthesis and fine-tuning a ResViT-based classifier model for brain tumor identification. In addition, the synthesized MRI images by ResViT have been included to enhance classification performance. The proposed model has various important components. Firstly, the self-supervised learning strategy allows for the learning distribution of MRI datasets during MRI synthesis in an unsupervised manner without tumor-type labels. Secondly, The ResViT-based model enables the utilization of Residual CNN and ViT together to gather local and global features from MRI datasets. Moreover, fine-tuning and data augmentation via synthetic MRI strategies enable a more effective and data-efficient approach to overcome overfitting on a small number of brain tumor datasets. In summary, the proposed model combines various strategies, self-supervised learning, hybrid architecture with CNN and ViT, fine-tuning, and data augmentation for a more effective and data-efficient approach.

The proposed model compares state-of-the-art models for MRI synthesis and classification, fine-tuning models, sequence types, and datasets. The proposed model performs evidently better than other state-of-the-art models for each dataset and each sequence. Implementing fine-tuning via a pre-trained model on an MRI dataset in the pretext step and synthetic MRI image has increased the accuracy of the proposed model rather than pre-trained models on ImageNet. The results show that T1 is a more convenient sequence to identify tumor types. Furthermore, the pretrained proposed model can easily transferred to other MRI datasets for fine-tuning. As a result, the proposed model emerges as a powerful approach to diagnosing brain tumors by providing high performance, flexibility, robustness, adaptability to diverse datasets, and strong generalization ability.

\section*{Acknowledgment}
This work was supported by The Scientific and Technological Research Council of Turkey (TUBITAK-BIDEB 2214/A) under project number 1059B142201736.The first author would like to thank The Scientific and Technological Research Council of Turkey (TUBITAK) and Biocomplexity Institute and Initiative, University of Virginia.

\end{document}